\documentclass[prl,twocolumn,superscriptaddress]{revtex4}
\usepackage{}%
\usepackage{amsfonts}
\usepackage{amsmath}
\usepackage{amssymb}
\usepackage{graphicx}%
\usepackage{color}
\emergencystretch=\maxdimen
\begin{document}

\title{Spin-Lattice Coupling Induced Rich Magnetic States in CrF$_3$ monolayer}

\author{Fawei Zheng}
\thanks{Corresponding author: fwzheng@bit.edu.cn}
\affiliation{Centre for Quantum Physics, Key Laboratory of Advanced Optoelectronic Quantum Architecture and Measurement (MOE), School of Physics, Beijing Institute of Technology, Beijing 100081, China.}
\affiliation{Beijing Key Lab of Nanophotonics $\&$ Ultrafine Optoelectronic Systems, School of Physics, Beijing Institute of Technology, Beijing 100081, China.}

\author{Yong Lu}
\thanks{Corresponding author: luy@mail.buct.edu.cn}
\affiliation{College of Mathematics and Physics, Beijing University of Chemical Technology, Beijing 100029, China. }


\date{\today}
\clearpage

\begin{abstract}
We systematically studied the spin-lattice couplings in the CrF$_3$ monolayer. Our study reveals that the spin exchange constants between the nearest neighbors are notably affected by these couplings. Specifically, the couplings arise predominantly from three distinct phonon modes, namely the covariant, rotation, and stretch of the Cr-F-Cr-F rhombus. By integrating out the phonon degrees of freedom, we derived an effective spin Hamiltonian featuring four-spin product terms, which yields a remarkably intricate magnetic phase diagram. Significantly, numerous plateau states characterized by fractional magnetizations, including 1/2, 1/3, 2/3, 1/4, 1/5, 5/8, 1/9, and 2/9, emerge in the vicinity of the phase transition boundary separating ferromagnetic and antiferromagnetic states. These findings show the profound influence of spin-lattice couplings on magnetic properties near the magnetic phase boundaries, and the predicted plateau states are expected to be observable in future experiments.
\end{abstract}

\maketitle
\clearpage
The spin-lattice coupling (SLC) has been studied for almost a century \cite{Heisenberg1928, Zener1951,Anderson1955, Penc2004, Fennie2006, Bergman2006, Kocsis2023}. It controls the spin relaxation rate and produces a variety of interesting phenomena, such as topological magnon polarons \cite{Go2019,Bao2023}, phonon-magneto chiral effects \cite{Thingstad2019,Nomura2019}, strong ferroelectric ferromagnets \cite{Petit2007,Cheong2007}, spin-lattice liquids \cite{Isono2018,Smerald2019}, spin-Peierls instability \cite{Becca2002,Routh2022}, and angular momentum transport between spin and lattice \cite{Strugaru2021,Mankovsky2022}. Compared to bulk systems, the SLC effect is more significant in two-dimensional (2D) magnetic systems, which has been confirmed in 2D Fe$_3$GeTe$_2$ and FePS$_3$ via magneto-Raman spectroscopy and comprehensive analysis of anomalous
phonon energy and linewidth \cite{Du2019,Liu2021,Cai2023}. The SLC effect has also been studied in 2D Cr$X$$_3$ ($X$=Cl, Br, I) \cite{Huang2017,Pocs2020,Kozlenko2021} and Cr$Y$Te$_3$ ($Y$=Si, Ge) \cite{Gong2017,Milosavljevic2018}. Theoretical calculations demonstrate that the SLC effect in monolayer CrI$_3$ is up to ten times larger than that of bcc Fe \cite{Sadhukhan2022,Hellsvik2019}.

The SLC has a prominent impact in 2D spin-frustrated systems. For example, the existence of a half-magnetization plateau and a uniform XY magnetization in antiferromagnetic (AFM) pyrochlore lattice have been proven as the results of SLC \cite{Hagymasi2022, Gen2023}. More complex collinear orders, such as 1/5, 1/3, 3/7, 3/5, and 5/7, have also been predicted in AFM triangular and kagome lattices, elucidating the collinear states in these magnetic structures \cite{CuFeO2-1,CuFeO2-2,NaFeO2-1,NaFeO2-2,Gen2022}.

\begin{figure}
	\centering
	\includegraphics[width=0.9\columnwidth]{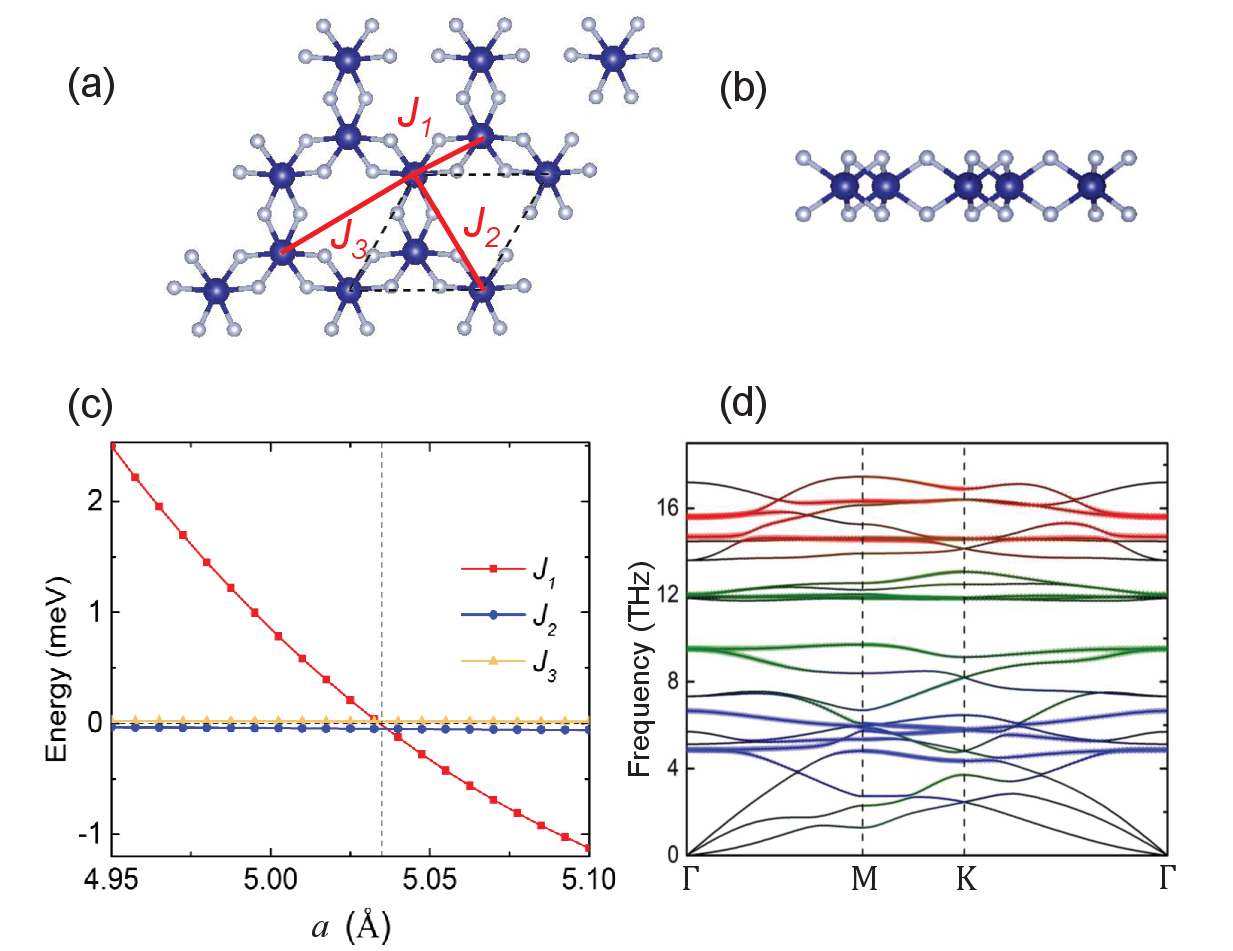}
	\caption{(Online color) (a) Top view and (b) side view of the CrF$_3$ monolayer. (c) The exchange constants of the first ($J_1$), second ($J_2$), and third ($J_3$) nearest neighbors (in units of meV) as a function of the lattice constant $a$ (\AA). (d) The phonon dispersion curves of CrF$_3$ monolayer. The green color represents the projection of the covariant mode, while the red and blue colors represent the projection of the stretching and rotating modes of the Cr-F-Cr-F rhombus. The blue and white balls in (a) represent the Cr and F atoms, respectively.
		\label{fig:Graph1}}
\end{figure}

In this work, we propose that the SLC has a significant impact on the phase transition point between ferromagnetism (FM) and AFM in CrF$_3$ monolayer, leading to the emergence of multiple new phases. Based on model parameters obtained from density functional theory (DFT) calculations, we demonstrate that only nearest-neighbor spin interactions exhibit a strong SLC effect, with three different types of phonons contributing to a major part of the SLC.
The contribution of other phonons or SLC in spin interactions beyond the nearest neighbors is negligible. We find that the introduction of SLC significantly alters the magnetic phase diagram. Interestingly, many collinear magnetic states emerge around the FM-AFM phase boundary. These new states form complex magnetization plateaus and can be detected by magnetic susceptibility measurements.

The CrF$_3$ monolayer has a structure similar to that of the CrI$_3$ monolayer. They are two members of the Cr$X_3$ family, where $X$ represents halogen atoms. The large magnetic moment of each Cr atom (3$\mu_B$) suppresses quantum fluctuations, thus a classical spin lattice model is applicable.  The unit cell of a Cr$X_3$ monolayer contains two Cr atoms and six $X$ atoms, as shown in Fig. 1(a). Six $X$ atoms surround a Cr atom, forming a slightly distorted octahedron. The magnetic Cr atoms form a graphene-like honeycomb lattice, where each adjacent Cr atoms are connected by a pair of $X$ atoms. From a side view, as shown in Fig. 1(b), the Cr atom layer is sandwiched between two $X$ atom layers. The monolayers of Cr$X_3$ exhibit similar magnetic phase diagrams. Specifically, they exhibit ferromagnetic (FM) states at large lattice constants and undergo a transition to antiferromagnetic (AFM) states at small lattice constants. This is the simplest SLC effect that arises from strain. Additionally, more nontrivial SLC effects can arise from phonon vibrations.

Normally, the variation of the spin exchange constant $J$ introduced by the SLC is quite small compared to its original value. However, around the magnetic phase transition point, the $J$ values are almost cancelled out by each other, in which case SLC may control the magnetic phenomena. It has been reported that the compressive strain required for the transition from FM to AFM in the monolayer CrI$_3$ is as high as 5.7 \% \cite{Zheng2018}. Our calculations indicate that the FM-AFM transition in the CrF$_3$ monolayer can occur at a much lower compressive strain of 3.1 \%. Since the CrF$_3$ monolayer does not contain heavy elements, the relativistic effect is negligible. Therefore, a simple Heisenberg model can effectively describe its magnetic properties. The simplicity of the model Hamiltonian and the easily accessible FM-AFM phase transition make the CrF$_3$ monolayer an ideal material for studying the SLC effect. Furthermore, this research can provide valuable insights into the SLC effect in other 2D magnetic materials.

\begin{figure}
\includegraphics[width=0.9\columnwidth]{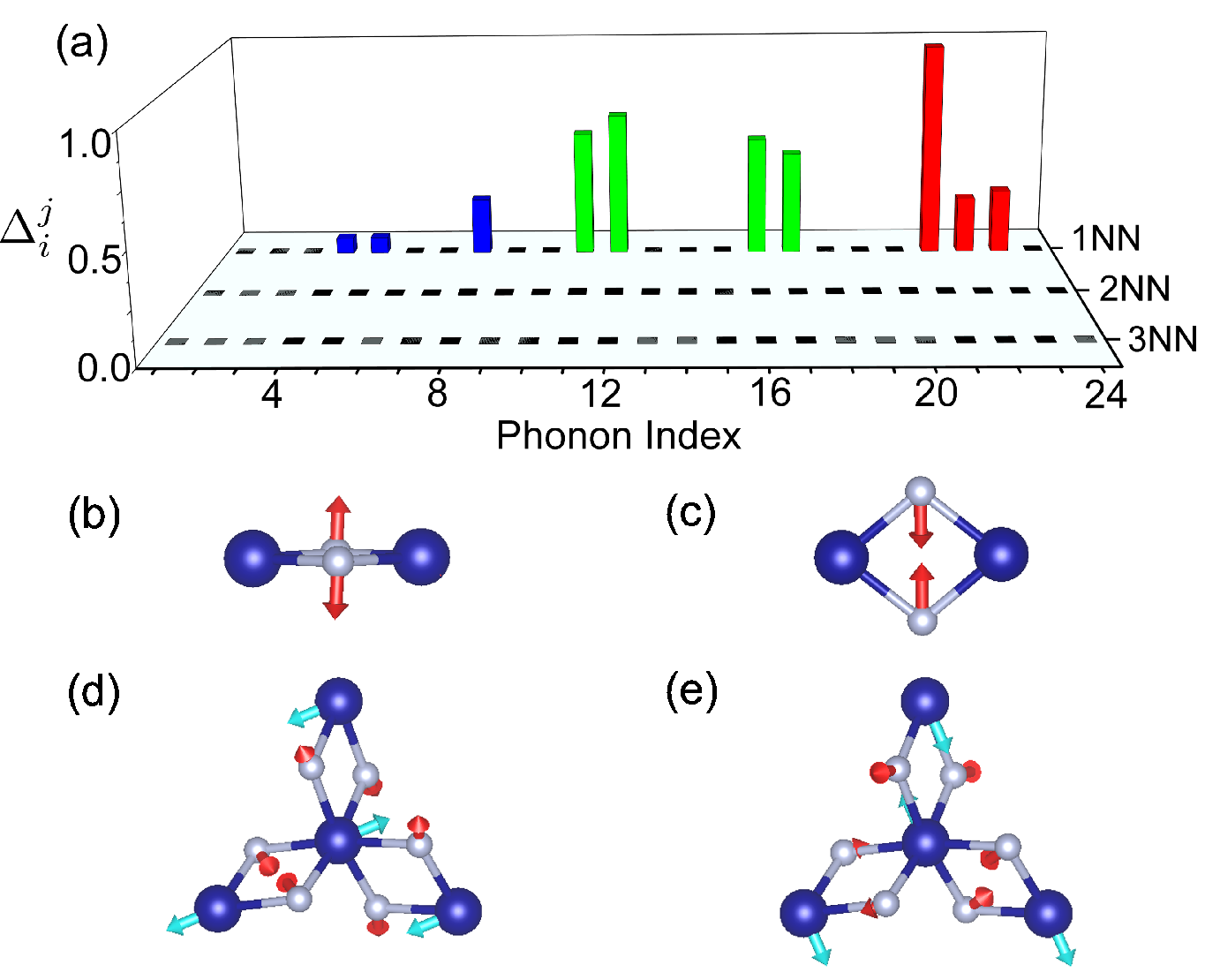}
\caption{(Color online) (a) The calculated $\Delta_{i}^j$ (meV) for the first (1NN), second (2NN), and third (3NN) nearest neighbors caused by all 24 phonon modes of CrF$_3$ monolayer. Different vibrational vectors of (b) Cr-F-Cr-F rhombus rotation mode, (c) Cr-F-Cr-F rhombus stretching mode, and (d)-(e) covariant mode are displayed.}
\end{figure}

The Heisenberg model describing the magnetic properties of the fully relaxed CrF$_3$ monolayer is
\begin{equation*}
\begin{aligned}
H_0=& \sum_{\langle ij \rangle}J_1 \vec{S}_{i}\cdot \vec{S}_{j}+\sum_{\langle\langle ij \rangle\rangle}J_2 \vec{S}_{i}\cdot \vec{S}_{j} +\sum_{\langle\langle\langle ij \rangle\rangle\rangle}J_3 \vec{S}_{i}\cdot \vec{S}_{j}\\&  +\vec{h}\cdot\sum_i\vec{S}_{i},
\end{aligned}
\end{equation*}
where the summation  $\langle ij\rangle$ operates on the nearest neighboring Cr atoms in the lattice, and $\langle\langle ij\rangle\rangle$ and  $\langle\langle\langle ij \rangle\rangle\rangle$ act on the next-nearest neighbors and third-nearest neighbors, respectively, representing their  distances.  The $\vec{S}_{i}$ and $\vec{S}_{j}$ are the angular momentums of the $i$-th and $j$-th sites, with a length of $\frac{3}{2}$. Here, $\vec{h}$ is proportional to the external magnetic field. The parameters $J_1$, $J_2$, and $J_3$ are obtained by energy mapping of four different magnetic states in a 2$\times$1 supercell, as shown in the Supplemental Material Fig. S1 \cite{Supp}. The calculated results are shown in Fig. 1(c), where the value of $J_1$ undergoes a significant change under strain and a sign inversion occurs at $a$=5.027 \AA. The values of $J_2$ and $J_3$ are -0.059 eV and 0.017 eV at 5.10 \AA, and slightly change to -0.034 eV and 0.022 eV at 4.95 \AA. Although their values are relatively small, they cannot be neglected near the phase transition point. Under the atomic distortion caused by phonon modes, the spin exchange interaction between Cr atoms may change. We denote the altered exchange parameter as $J_i^{jk}=J_i+\Delta_i^{jk}$ ($i$=1, 2, 3; $j$=1, 2, 3, ...; $k$=1, 2, 3 or 1, 2, 3, 4, 5, 6). The integer $i$ represents the shell of neighbors, $j$ is the phonon index, and $k$ defines the direction of the exchange interaction. $\Delta_i^{jk}$ denotes the change of $J_i$ in the $k$-th direction caused by the $j$-th phonon mode. The values of $\Delta_i^{jk}$ can be obtained by combining magnetic force theory calculation with frozen phonon method. We measure the overall change of the exchange constant caused by the $j$th phonon by  $\Delta_i^j=\sqrt{(\Delta_i^{j1})^2+(\Delta_i^{j2})^2+...+(\Delta_i^{j24})^2}$. The calculated values of $\Delta_i^j$ caused by all 24 phonon branches at the $\Gamma$ point are shown in Fig. 2(a). There are 10 phonon modes (P$_4$, P$_5$, P$_8$, P$_{11}$, P$_{12}$, P$_{16}$, P$_{17}$, P$_{21}$, P$_{22}$, and P$_{23}$) contribute significantly, while the values of $\Delta_i^j$ for the other 14 phonon modes are almost zero. The projection weights of these 10  phonon modes in the phonon dispersions are shown in Fig. 1(d).

By analyzing the detailed polarization vectors, we classify these 10 phonon modes into three categories. The corresponding strengths of SLC ($\Delta_i^j$) are shown in Fig. 2(a) by different colors. The phonon modes $P_4$, $P_5$ and $P_8$ involve vibrations of the F atoms perpendicular to the Cr-F-Cr-F rhombus. They are rotations of the rhombus, with the two F atoms within the same rhombus vibrating in the opposite directions while the Cr atom remains stationary, as shown in Fig. 2(b). There are three Cr-F-Cr-F rhombuses in a cell of CrF$_3$ monolayer, and their rotations are combined into the $P_4$, $P_5$, and $P_8$ modes. The rotation of the rhombus only changes the bond angle of F-Cr-F and Cr-F-Cr, without affecting the Cr-F bond length, resulting in relatively weak stiffness. This leads to low-frequency phonons at around 5 THz. We use the collective coordinate $u_{ij}$ to describe the rotation of the rhombus between $i$-th and $j$-th Cr atoms. Similarly, the $P_{21}$, $P_{22}$ and $P_{23}$ phonon modes are also vibrations of the F atoms in the Cr-F-Cr-F rhombus, but they are the in-plane stretching modes, as shown in Fig. 2(c). The stretching modes of the three different rhombuses in a cell is combined to form the three phonon modes. The stretching of a rhombus directly alters the F-Cr bond length, resulting in a stronger stiffness. This makes the three phonon modes have a high frequency of around 15 THz. The collective coordinates describing the stretching of the rhombuses between the $i$-th and $j$-th Cr atoms are denoted as $v_{ij}$. The Hamiltonian of these vibration modes and the SLC can be written as
\begin{equation}
H_1= \sum_{\langle ij \rangle}(\lambda_{1}u_{ij} +\lambda_{2}v_{ij})\vec{S}_{i}\cdot \vec{S}_{j} + \frac{1}{2}\sum_{\langle ij \rangle}(k_{1}u_{ij}^{2} +k_{2}v_{ij}^{2}),
\end{equation}
where $\lambda_{1}$ and $\lambda_{2}$ are the SLC constants caused by the rotation and stretching of the Cr-F-Cr-F rhombus, while $k_{1}$ and $k_{2}$ are the stiffness constants of these vibrations. This equation is similar to the bond-phonon (BP) model \cite{Penc2004}. The difference is that the bond displacements here is not the change in Cr-Cr bond length, but the change in F atoms between each two neighboring Cr atoms, which is appropriate to describe the present system. The parameters in $H_1$ can be obtained by using the total energy calculation of the primitive cell. By adding different amounts of rotation ($u_{ij}$) step by step to the Cr-F-Cr-F rhombuses, the stiffness parameter $k_{1}$ can be obtained by using the total energy change. In each step, the exchange energy, \emph{i.e.}, the energy difference between the FM state and the AFM state, is proportional to $H_0$ and the first term of $H_1$. Since $H_1$ is linearly dependent on $u_{ij}$ while $H_0$ is independent of $u_{ij}$, the parameter $\lambda_{1}$ can be obtained from the slope of the exchange energy. Similarly, we can also obtain the values of $k_2$ and $\lambda_2$.

\begin{figure}
	\centering
	\includegraphics[width=0.9\columnwidth]{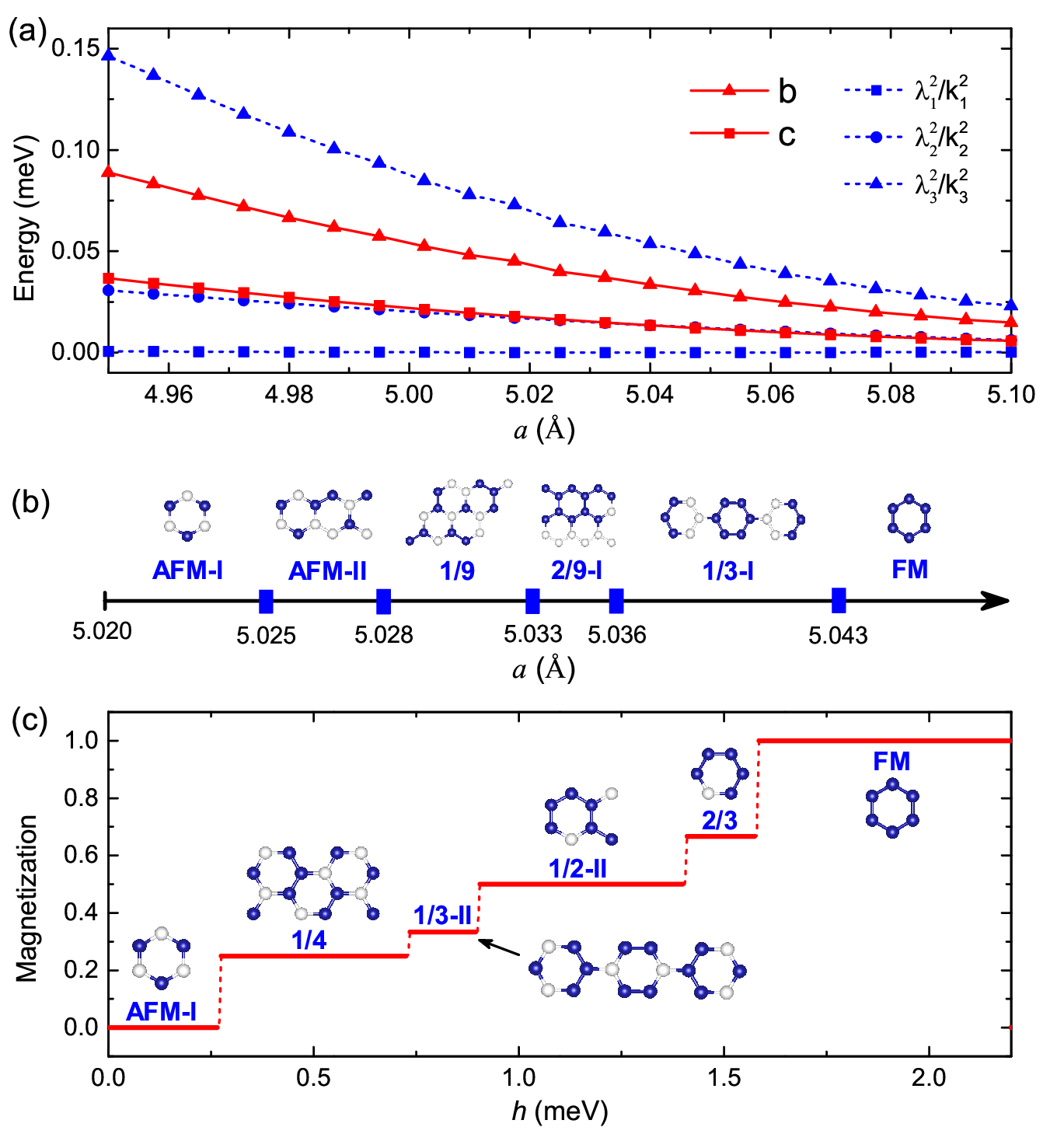}
	\caption{(Online color) (a) The calculated values of $\lambda_i^2/k_i^2$, $b$ and $c$ in $H_{eff}$ (in units of meV) as a function of the lattice constant $a$ (\AA). (b) Magnetic phases of monolayer CrF$_3$ at different lattice constants under zero magnetic field. (c) The predicted magnetization curve as a function of applied field $h$ (meV) at $a$=5.025 \AA. Solid blue (white) balls represent sites with up (down) spin.
		\label{fig:Graph1}}
\end{figure}

The remaining four vibration modes can be classified into two lower-frequency degenerate modes ($P_{11}$ and $P_{12}$) and two higher-frequency degenerate modes ($P_{16}$ and $P_{17}$ ). These modes involve the vibrations of both F and Cr atoms and exhibit more complex behavior. In order to get vibration modes that are more crucial to $\Delta$, we performed singular value decomposition on the polarization vectors. Two independent vibrational modes were successfully extracted, as shown in Fig. 2(d)-(e).  The two Cr atoms vibrate in opposite directions, analogous to in-plane optical phonons in graphene. The surrounding F atoms move around the Cr atoms like jellium, moving according to the positions of the nearby Cr atoms. Specifically, they move outward (inward) when the two Cr atoms squeeze (stretch) the rhombus. We refer to this covariant moving of Cr and F atoms as a covariant mode and it can be described by the relative positions between the two Cr atoms in the rhombus. The Hamiltonian of these two vibration modes and the SLC can be written as
\begin{equation}
H_2= \sum_{\langle ij \rangle}\lambda_{3} \hat{e}_{ij}\cdot (\vec{r}_{i}-\vec{r}_{j})\vec{S}_{i}\cdot \vec{S}_{j} + \frac{1}{2}\sum_{i}k_{3}\vec{r}_{i}^{2},
\end{equation}
where $\lambda_{3}$ is the SLC constant, $k_{3}$ is the stiffness constant when stretching the rhombus between two Cr atoms, $\vec{r}_{i}$ and $\vec{r}_{j}$ are the displacement vectors at sites $i$ and $j$, and $\hat {e}_{ij}$ is the normalized vector from site $i$ to $j$. This is an Einstein-site-phonon (ESP) model \cite{Bergman2006}, which describes the SLC of independent displacement of each magnetic atom. The parameters in $H_2$ can also be obtained from the total energy calculations of the primitive cell.
By applying biaxial strain to the system step by step, one Cr atom is fixed at the center, while the other Cr atom moves closer to it gradually. The change in total energy is proportional to the second term in $H_2$, allowing us to obtain the value of $k_3$. Then, we calculated the exchange energy and obtained the value $\lambda_3$ through its slope.

The parameters $J$, $\lambda_1$, $\lambda_2$, $\lambda_3$, $k_1$, $k_2$, and $k_3$ in the complete spin-lattice Hamiltonian $H=H_0+H_1+H_2$ are tunable by biaxial strain. Using the same technique described in the previous works by Bergman \cite{Bergman2006} and Wang \cite{Wang2008}, the lattice degrees of freedom $r$, $u$, and $v$ can be integrated out, resulting in an effective spin Hamiltonian as,
\begin{equation}\label{Heff}
H_{eff}= H_0 - b\sum_{\langle ij \rangle} (\vec{S}_{i}\cdot \vec{S}_{j})^{2} +c\sum_{\langle ij \rangle \langle jk \rangle \atop j\neq k} (\vec{S}_{i}\cdot \vec{S}_{j}) (\vec{S}_{i}\cdot \vec{S}_{k}),
\end{equation}
where $b=\frac{\lambda_{1}^{2}}{2k_{1}} + \frac{\lambda_{2}^{2}}{2k_{2}} +\frac{\lambda_{3}^{2}}{2k_{3}}$ and $c=\frac{\lambda_{3}^{2}}{4k_{3}}$. This effective spin Hamiltonian contains the biquadratic interaction terms $(\vec{S}_{i}\cdot \vec{S}_{j})^{2}$ originating from both $H_1$ and $H_2$, and the three-neighborhood spin interaction terms $(\vec{S}_{i}\cdot \vec{S}_{j}) (\vec{S}_{i}\cdot \vec{S}_{k})$ originating from $H_2$. The parameters $b$ and $c$ are also functions of lattice constants. The values of $\frac{\lambda_i^2}{k_i^2}$ are shown in Fig. 3(a), from which it can be seen that the contribution from the covariant modes is the largest, while that from the rhombus rotation modes is negligible. As the lattice constant approaches the FM-AFM phase transition point, the value of $J$ vanishes gradually, while the values of $b$ and $c$ remain finite. Then, these high-order spin interaction terms control the magnetic properties of the system.

To obtain the ground magnetic state with various lattice constants and magnetic fields, we employ a combination of computations using the Genetic Algorithm (GA) and the Landau-Lifshitz-Gilbert (LLG) equation \cite{LLG1,LLG2}. The GA method is a general optimization technique inspired by biological evolution, which has recently been introduced for predicting magnetic structures \cite{Zheng2021}. In this work, we apply the GA method to generate and screen new magnetic structures, which were optimized by using the LLG equation. The flowchart of the prediction process is shown in the Supplemental Material Fig. S2 \cite{Supp}.

Fig. 3(b) shows the phase diagram at zero magnetic field. The N\'eel-AFM state (AFM-I) exists at small lattice constants ($<$ 5.025 \AA). Another AFM state (AFM-II) exists between 5.025 \AA$~$ and 5.028 \AA, which contains a magnetic unit with 12 magnetic atoms. On the other hand, the FM state only appears for large lattice constants ($>$ 5.043 \AA). Between the AFM-II and FM states, there are plateau states at 1/9, 2/9-I and 1/3-I. As the lattice constant increases, the magnetization increases from 0 to 1 stage by stage.  By increasing the magnetic field, the magnetic properties change in a similar trend. As shown in Fig. 3(c), the AFM-I state exists at weak magnetic field, while the FM state exists at strong magnetic field. There are plateau states between the AFM-I and FM states at 1/4, 1/3-II, 1/2-II, and 2/3. Sudden changes in magnetization occur between these states, which can be detected in experiments.

The numerically obtained phase diagram of the CrF$_3$ monolayer is shown in Fig. 4. If the SLC is ignored, the phase diagram is very simple, which contains the AFM-I state for small lattice constants and the FM state for large lattice contants, as divided by the black dashed line in Fig. 4. However, after considering the SLC, the phase diagram becomes significantly richer. There emerge the AFM-II state and fractional magnetic plateaus of 1/9, 1/5, 2/9-I, 2/9-II, 1/4, 3/11, 1/3-I, 1/3-II, 1/2-I, 1/2-II, 5/8, and 2/3. These new phases exist near the FM-AFM boundary. By increasing the lattice constant, the spin exchange parameter $J_1$ changes from AFM to FM, and the spins tend to have a parallel arrangement. In turn, the plateaus tend to have larger fractional magnetic moments, eventually becoming the FM state for large lattice constants. On the other hand, by enhancing the strength of the magnetic field, the plateau also tend to have larger fractional magnetic moments, as the magnetic field also favors a parallel spin arrangement. It is worth noting that the AFM state and several fractional magnetic states, such as  2/9, 1/3, and 1/2, have multiple spin configurations. The detailed information about these spin configurations can be found in the Supplemental Material Fig. S3 \cite{Supp}.

\begin{figure}
\centering
\includegraphics[width=1.0\columnwidth]{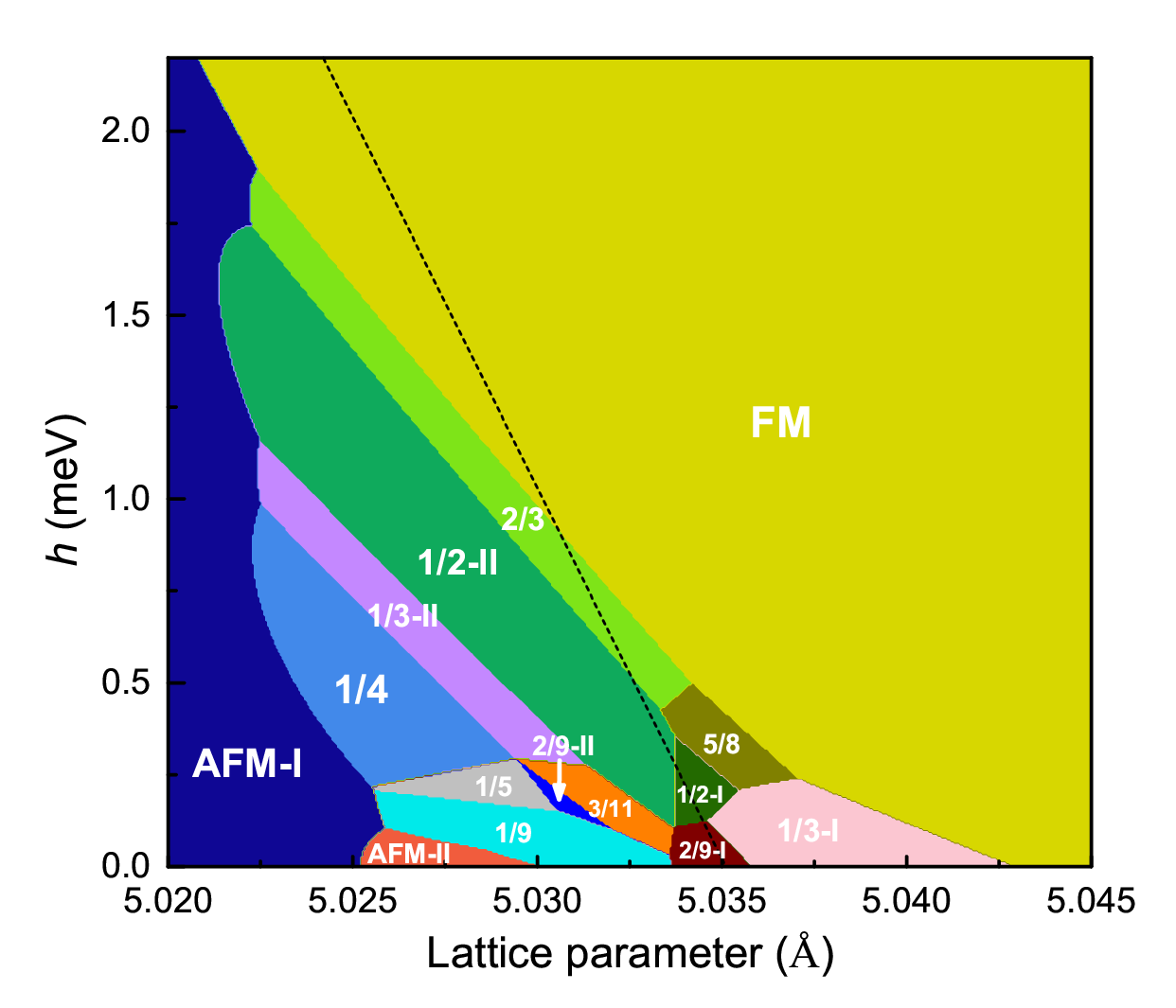}
\caption{(Online color) Phase diagram of monolayer CrF$_3$ under different lattice constants and magnetic fields. Characters or fractions label different magnetic plateaus. The black dashed line represents the boundary between the FM phase and the AFM phase without considering the SLC case.
\label{fig:Graph1}}
\end{figure}

In summary, by using magnetic force theory calculations and frozen phonon method, we have found that the SLC in the CrF$_3$ monolayer is mainly contributed by three different types of phonon modes, namely the covariant, rotational, and stretching modes of the Cr-F-Cr-F rhombus. The BP-ESP-like models were constructed, and an effective spin Hamiltonian was obtained by integrating out the phonon degrees of freedom. By using the GA method and the LLG equation, we obtained the magnetic phase diagram of the CrF$_3$ monolayer. Compared to the phase diagram neglecting SLC, it is much more complex, with many plateau states emerging around the FM-AFM phase boundary. Our study demonstrates that SLC plays a crucial role in the magnetic properties near the phase transition point. This phenomenon may also exist in other magnetic materials. Experimental investigations are expected to confirm this prediction.

This work was supported by the National Natural Science Foundation of China under Grant Nos. 12022415, 11974056, 12074028, and 12374054. We acknowlege the computing resources of Tencent TEFS platform (https://tefscloud.com).


\begin{thebibliography}{35}
\bibitem{Heisenberg1928}
W. Heisenberg, Z. Physik {\bf 49}, 619-636 (1928).

\bibitem{Zener1951}
Clarence Zener, Phys. Rev. {\bf 82}, 403 (1951).

\bibitem{Anderson1955}
P. W. Anderson and H. Hasegawa, Phys. Rev. {\bf 100}, 675 (1955).

\bibitem{Penc2004}
K. Penc, N. Shannon, and H. Shiba, Phys. Rev. Lett. {\bf 93}, 197203 (2004).

\bibitem{Fennie2006}
C. J. Fennie and K. M. Rabe, Phys. Rev. Lett. {\bf 96}, 205505 (2006).

\bibitem{Bergman2006}
D. L. Bergman, R. Shindou, G. A. Fiete, and L. Balents, Phys. Rev. B {\bf 74}, 134409 (2006).


\bibitem{Kocsis2023}
V. Kocsis, Y. Tokunaga, T. R\~o\~om, U. Nagel, J. Fujioka, Y. Taguchi, Y. Tokura, and S. Bord\'acs, Phys. Rev. Lett. {\bf 130}, 036801 (2023).

\bibitem{Go2019}
G. Go, S. K. Kim, and K. J. Lee, Phys. Rev. Lett. {\bf 123}, 237207 (2019).

\bibitem{Bao2023}
S. Bao, Z. Gu, Y. Shangguan, Z. Huang, J. Liao, X. Zhao, B. Zhang, Z. Dong, W. Wang. R. Kajimoto, M. Nakamura, T. Fennell, S. Yu, J. Li, J. Wen, Nat. Commun. {\bf 14}, 6093 (2023).

\bibitem{Thingstad2019}
E. Thingstad, A. Kamra, A. Brataas, and A. Sudb\o, Phys. Rev. Lett. {\bf 122}, 107201 (2019).

\bibitem{Nomura2019}
T. Nomura, X.-X. Zhang, S. Zherlitsyn, J. Wosnitza, Y. Tokura, N. Nagaosa, and S. Seki, Phys. Rev. Lett. {\bf 122}, 145901 (2019).

\bibitem{Petit2007}
S. Petit, F. Moussa, M. Hennion, S. Pailh\'es, L. Pinsard-Gaudart, and A. Ivanov, Phys. Rev. Lett. {\bf 99}, 266604 (2007).

\bibitem{Cheong2007}
S. W. Cheong and M. Mostovoy, Nat. Mater. {\bf 6}, 13 (2007).

\bibitem{Isono2018}
T. Isono, S. Sugiura, T. Terashima, K. Miyagawa, K. Kanoda, S. Uji, Nat. Commun. {\bf 9}, 1509 (2018).

\bibitem{Smerald2019}
A. Smerald and G. Jackeli, Phys. Rev. Lett. {\bf 122}, 227202 (2019).

\bibitem{Routh2022}
M. Routh, S. K. Saha, M. Kumar, and Z. G. Soos, Phys. Rev. B {\bf 105}, 235109 (2022).

\bibitem{Becca2002}
F. Becca, F. Mila, Phys. Rev. Lett. {\bf 89}, 037204 (2002).

\bibitem{Strugaru2021}
M. Strungaru, M. O. A. Ellis, S. Ruta, O. C.- Fesenko, R. F. L. Evans, and R. W. Chantrell, Phys. Rev. B {\bf 103}, 024429 (2021).

\bibitem{Mankovsky2022}
S. Mankovsky, S. Polesya, H. Lange, M. Wei$"s$enhofer, U. Nowak, and H. Ebert, Phys. Rev. Lett. {\bf 129}, 067202 (2022).

\bibitem{Du2019}
L. Du, J. Tang, Y. Zhao, X. Li, R. Yang, X. Hu, X. Bai, X. Wang, K. Watanabe, T. Taniguchi, D. Shi, G. Yu, X. Bai, T. Hasan, G. Zhang, and Z. Sun, Adv. Funct. Mater. 1904734 (2019).

\bibitem{Liu2021}
S. Liu, A. G. \'Aguila, D. Bhowmick, C. K. Gan, T. T. H. Do, M. A. Prosnikov, D. Sedmidubsk\'y, Z. Sofer, P. C. M. Christianen, P. Sengupta, and Q. Xiong, Phys. Rev. Lett. {\bf 127}, 097401 (2021).

\bibitem{Cai2023}
Q. Cai, Y. Zhang, D. Luong, C. A. Tulk, B. P.T. Fokwa,and C. Li, Adv. Physics Res. {\bf 2}, 2200089 (2023).


\bibitem{Huang2017}
B. Huang, G. Clark, E. Navarro-Moratalla, D. R. Klein, R. Cheng, K. L. Seyler, D. Zhong, E. Schmidgall, M. A. Mcguire, D. H. Cobden, W. Yao, D. Xiao, P. Jarillo-Herrero, X. Xu, Nature {\bf 546}, 270 (2017).


\bibitem{Pocs2020}
C. A. Pocs, I. A. Leahy, H. Zheng, G. Cao, E. Choi, S.-H. Do, K. Choi, B. Normand, and M. Lee, Phys. Rev. Res. {\bf 2}, 013059 (2020).

\bibitem{Kozlenko2021}
D. P. Kozlenko, O. N. Lis, S. E. Kichanov, E. V. Lukin, N. M. Belozerova and B. N. Savenko, npj Quantum Mater. {\bf 6}, 19 (2021).

\bibitem{Gong2017}
C. Gong, L. Li, Z. Li, H. Ji, A. Stern, Y. Xia, T. Cao, W. Bao, C. Wang, Y. Wang, Z. Q. Qiu, R. J. Cava, S. G. Louie, J. Xia, X. Zhang, Nature
{\bf 546}, 265 (2017).

\bibitem{Milosavljevic2018}
A. Milosavljevi\'c, A. \v Solaji\'c, J. Pe\v si\'c, Y. Liu, C. Petrovic, N. Lazarevi\'c, and Z. V. Popovi\'c, Phys. Rev. B {\bf 98}, 104306 (2018).

\bibitem{Sadhukhan2022}
B. Sadhukhan, A. Bergman, Y. O. Kvashnin, J. Hellsvik, and A. Delin, Phys. Rev. B {\bf 105}, 104418 (2022).

\bibitem{Hellsvik2019}
J. Hellsvik, D. Thonig, K. Modin, D. Iu?san, A. Bergman, O. Eriksson, L. Bergqvist, and A. Delin, Phys. Rev. B {\bf 99}, 104302 (2019).

\bibitem{Hagymasi2022}
I. Hagym\'asi, R. Sch\"afer, R. Moessner, and D. J. Luitz, Phys. Rev. B {\bf 106}, L060411 (2022)

\bibitem{Gen2023}
M. Gen, A. Ikeda, K. Aoyama, Harald O. Jeschke, Y. Ishii, H. Ishikawa, T. Yajima, Y. Okamoto,  X. Zhou, D. Nakamura, S. Takeyama, K. Kindo, Y. H. Matsuda, and Y. Kohama, PNAS {\bf 120}, No.33, e2302756120 (2023).

\bibitem{CuFeO2-1}
S. Mitsuda, M. Mase, K. Prokes, H. Kitazawa, and H. A. Katori, J. Phys. Soc. Jpn. {\bf 69}, 3513 (2000).

\bibitem{CuFeO2-2}
F. Ye, Y. Ren, Q. Huang, J. A. Fernandez-Baca, P. Dai, J. W. Lynn, and T. Kimura, Phys. Rev. B {\bf 73}, 220404(R) (2006).

\bibitem{NaFeO2-1}
T. McQueen, Q. Huang, J. W. Lynn, R. F. Berger, T. Klimczuk, B. G. Ueland, P. Schiffer, and R. J. Cava, Phys. Rev. B {\bf 76}, 024420 (2007).

\bibitem{NaFeO2-2}
N. Terada, Y. Ikedo, H. Sato, D. D. Khalyavin, P. Manuel, A. Miyake, A. Matsuo, M. Tokunaga, and K. Kindo, Phys. Rev. B {\bf 96}, 035128 (2017).

\bibitem{Gen2022}
M. Gen and H. Suwa, Phys. Rev. B {\bf 105}, 174424 (2022).

\bibitem{Zheng2018}
F. W. Zheng, J. Zhao, Z. Liu, M. Li, M. Zhou, S. Zhang, and P. Zhang, Nanoscale {\bf 10}, 14298 (2018).

\bibitem{Supp}
See the Supplemental Material for additional information on the details of the DFT calculation, the calculation of spin exchange constants, the flowchart of the ground magnetic state prediction process, and the spin structures of the ground states in the phase diagram.

\bibitem{Wang2008}
F. Wang and A. Vishwanath, Phys. Rev. Lett. {\bf 100}, 077201 (2008).

\bibitem{LLG1}
L. D. Landau and E. M. Lifshitz, Phys. Z. Sowjet. {\bf 8}, 153 (1935).

\bibitem{LLG2}
T. L Gilbert, IEEE Trans. Magn. {\bf 40}, 3443 (2004).

\bibitem{Zheng2021}
F. W. Zheng, P. Zhang, Comput. Phys. Commun. {\bf 259}, 107659 (2021).



\end{thebibliography}
\end{document}